\documentclass[lettersize,journal]{IEEEtran}
\usepackage{amsmath,amsfonts}
\usepackage{algorithmic}
\usepackage{algorithm}
\usepackage{array}
\usepackage[caption=false,font=normalsize,labelfont=sf,textfont=sf]{subfig}
\usepackage{textcomp}
\usepackage{stfloats}
\usepackage{url}
\usepackage{verbatim}
\usepackage{graphicx}
\usepackage{cite}
\usepackage{xcolor}

\DeclareMathOperator*{\sign}{\mathrm{sign}}

\usepackage{color,soul}
\soulregister\hl{7}
\soulregister\ref7
\soulregister\eqref7
\soulregister\cite7
\soulregister\pageref7
\definecolor{yellowhl}{rgb}{0.9,0.8,0.5}

\usepackage{tikz}
\newcommand\copyrighttext{%
  \centering\footnotesize \textbf{\copyright 2023 IEEE}.  Personal use of this material is permitted.  Permission from IEEE must be obtained for all other uses, in any current or future media, including reprinting/republishing this material for advertising or promotional purposes, creating new collective works, for resale or redistribution to servers or lists, or reuse of any copyrighted component of this work in other works. \\
  \textbf{Cite as:} S. Radrizzani, G. Panzani, L. Trezza, S. Pizzocaro, S. M. Savaresi, ``An Add-on Model Predictive Control Strategy for the Energy Management of Hybrid Electric Tractors", \textit{IEEE Transactions on Vehicular Technology}, 2023, doi: 10.1109/TVT.2023.3318237}
\newcommand\copyrightnotice{%
\begin{tikzpicture}[remember picture,overlay]
\node[anchor=south,yshift=-2cm] at (current page.north) {\fbox{\parbox{\dimexpr0.9\textwidth-\fboxsep-\fboxrule\relax}{\copyrighttext}}};
\end{tikzpicture}%
}

\begin{document}

\title{An Add-on Model Predictive Control Strategy for the Energy Management of Hybrid Electric Tractors}
\author{Stefano Radrizzani, Giulio Panzani, Luca Trezza, Solomon Pizzocaro, Sergio M. Savaresi
\thanks{All the authors are with the Dipartimento di Elettronica, Informazione e Bioingegneria, Politecnico di Milano, Via Ponzio 34/5, 20133 Milan, Italy. (e-mail: giulio.panzani@polimi.it).}}



\maketitle
\copyrightnotice
\begin{abstract}
The hybridization process has recently touched also the world of agricultural vehicles. Within this context, we develop an Energy Management Strategy (EMS) aiming at optimizing fuel consumption, while maintaining the battery state of charge. A typical feature of agricultural machines is that their internal combustion engine is speed controlled, tracking the reference requested by the driver. In view of avoiding any modification on this original control loop, an add-on EMS strategy is proposed. In particular, we employ a multi-objective Model Predictive Control (MPC), taking into account the fuel consumption minimization and the speed tracking requirement, including the engine speed controller in the predictive model. The proposed MPC is tested in an experimentally-validated simulation environment, representative of an orchard vineyard tractor.
\end{abstract}


\section{Introduction}
\IEEEPARstart{T}{he} hybridization process is currently moving from standard vehicles to heavy-duty ones, raising new challenges not only in their mechanical/electrical design but also in the development of a proper Energy Management Strategy (EMS).

Heavy-duty vehicles cover a wide and heterogeneous class of vehicles \cite{beltrami2021offroad}, including but not limited to agricultural tractors, street sweepers, forklifts and construction vehicles. Therefore, they cannot be studied as a single category, in order to exploit their potential. It follows that if, on the one hand, traditional vehicle behavior is well represented by standard driving-cycles, on the other hand, each heavy-duty vehicle is characterized by particular features. To provide some examples, tractors can operate in a transport scenario, when used for moving people or loads through trailers, or in a working scenario, where they operate the same machinery in field for hours; on the contrary, construction vehicles operate in a very different way, moving loads for hours in a stationary way. Moreover, heavy-duty vehicles are typically equipped with additional loads with respect to traditional vehicles. These loads can be mechanical, connected to the propulsion system through a power take-off (PTO), or hydraulic, connected through a system of pumps that feeds the hydraulic circuit.
This work focuses on tractors, in particular on a hybrid orchard vineyard tractor prototype. It is a \underline{non} plug-in parallel hybrid vehicle, i.e., an electric motor (EM) is mounted in parallel to the internal combustion engine (ICE) and the battery state of charge (SoC) is maintained by the engine itself.

Generally, hybrid electric vehicles (HEVs) need an energy management strategy: a control law that optimizes the usage of the multiple power sources, aiming at optimizing fuel consumption \cite{paganelli2002ecms}, efficiency \cite{radrizzani2022emp} emissions \cite{nuesch2014nox} or costs in an economical framework \cite{pozzato2020economic,pozzato2020leastclstly}. This optimization problem can be seen as a global optimization \cite{sciarretta2004optimal,kessels2008online}, in fact the optimization horizon coincides with the entire life of the vehicle. Given the impossibility to know a-priori the behavior of the vehicle, many real-time solutions have been developed, here classified according to their level of optimality. The simplest solutions are based on heuristic approaches \cite{baumann2000rulebased}, followed by the well-known Equivalent Consumption Minimization Strategy (ECMS) that reduces the global optimization problem to a local one by minimizing an equivalent consumption at any time instant \cite{paganelli2002ecms}.
The optimization problem can be also solved over a finite prediction horizon, applying Model Predictive Control (MPC) techniques \cite{borhan2012mpc,ripaccioli2009hybridmpc,balaji2009mpc}. 
Among their advantages, there is the possibility to turn them into multi-objective problems; indeed, additional aspects can be easily included in the predictive model, e.g., models specific for the engine turn-on phases \cite{yan2012transients}, battery thermal \cite{hu2020thermal} or aging dynamics \cite{cheng2019battery}.

In the agricultural sector, besides the advantages shared with traditional hybrid vehicles, a hybrid tractor can reduce the deposit of pollutants on crops and the emissions near workers \cite{gonzalez2016pollution}. Moreover, when the full-electric mode is activated, turning off the engine, it is possible to work indoor in greenhouses, completely eliminating the emissions. Nevertheless, hybridization and full-electrification of tractors are very challenging. First of all, because of the high power request to the battery for machining and the high energy for long full-electric cycles, the current technology is able to satisfy these requirements mainly for compact or orchard tractors \cite{caban2018market}. However, the electrification process is a general trend in agriculture and mature for small and light agricultural mobile robots, which require less energy and power. For this specific kind of vehicles, different energy storage systems have been developed and analyzed, ranging from batteries to hybrid fuel-cell and photovoltaic power propulsion systems, e.g.,\cite{gil2023agrirobots,ghobadpour2023agrirobot}.

\subsection*{Main topic and related works}
This work addresses the design and analysis of energy management strategies for the hybrid orchard/vineyard tractor prototype, in order to understand the potential fuel saving improvements with respect to its standard configuration which features the sole internal combustion engine. A peculiar characteristic of agricultural machines (and more generally heavy-duty vehicles) is that their ICE is already equipped with a speed controller \cite{lee2019engine,mocera2021model} that helps to ease the driver in performing their typical long and repetitive operations. 
	
Our work explicitly considers the presence of the built-in speed controller and proposes a control law that results as an \textit{add-on} to the original vehicle, i.e., does not require any change/intervention on the built-in speed controller. This is a desirable feature, that facilitates the development and transition to hybrid vehicles at an industrial level, also in a consolidated market like the agricultural one.

This specific topic, to the best of our knowledge, has not been yet addressed in the literature. One can, indeed, divide the related scientific background into two groups: 1) works dealing with the integrated energy management and speed control in traditional parallel hybrid electric vehicles (HEVs), without any built-in engine speed controller; 2) works oriented to the design and control of hybrid tractors.

The speed control and energy management problem, in traditional parallel HEVs, is typically handled by splitting the problem into two hierarchical and independent sub-problems: the high-level speed controller computes the total power to drive the vehicle at the reference speed; the lower-level controller, that implements the actual energy management strategy, splits the total requested power between the electric motor and the internal combustion engine, aiming at minimizing consumption. Some recent examples follow: \cite{kural2015emsacc} proposes an external MPC-based adaptive cruise controller (ACC) which computes the load torque and its prediction; then they are both used as input for an ECMS-based energy management. Also \cite{he2020emsacc} develops a hierarchical control law with an external cooperative adaptive cruise control and an internal heuristic energy management strategy. The hierarchical framework proposed in \cite{liu2022emsacc} is composed of a neural network, which is leveraged in the upper layer controller to regulate the vehicle speed; then a genetic algorithm is exploited to determine the equivalence factor for the lower layer ECMS-based control. Finally, \cite{ruan2022emsacc} formulates a constrained finite-time optimization control problem, realized by two hierarchical model predictive controllers. None of the discussed works can be directly employed on the considered tractor, because of the built-in speed controller. Its presence, in fact, causes two main issues. Firstly, the ICE itself is already responsible for the speed tracking, making impossible to address the fuel saving and the speed tracking objectives as two decoupled problems. Secondly, the total torque/power cannot be directly split between the electric motor and the engine, because in the latter it is autonomously regulated by the built-in controller.

Considering the literature on hybrid tractors, different kinds of works can be found. In \cite{troncon2020tractor}, the hybridization of a tractor is addressed as a feasibility study, showing the potential benefits of hybrid powertrains. Other works analyze, from an electric and mechanical perspective, specific hybrid configurations, e.g., parallel tractors with electric motors mounted on the main shaft \cite{mocera2020analysis} or directly in wheels \cite{baek2022inwheel}. Only a few works propose EMS for hybrid tractors: \cite{ghobadpour2021intelligent} and \cite{barthel2014ems} approach the EMS problem using a backward modeling paradigm, where the driving-cycle is known in advance and can be perfectly tracked; \cite{zhen2022adaptive}, on the other side, does consider the speed profile tracking problem but assumes 
the possibility of controlling both the ICE and the electric motor torque, as in traditional parallel hybrid electric vehicles.

\subsection*{Main contributions and outline}
In our work, we address the EM of a hybrid tractor accounting for its specificity, i.e. the built-in engine speed controller. To do so, we propose an MPC-based EM solution, that takes into account the engine controller dynamics within the predictive model and that addresses two main objectives: fuel saving and speed reference tracking.

A preliminary step is the identification of a control-oriented vehicle model, to be used inside the MPC formulation. The identification process is also oriented to the creation of a more complex model, to be used as simulation environment to test the proposed EMS. We mainly tested the proposed strategies on transport driving-cycles instead of agricultural ones. Indeed, agricultural operations typically occur at constant speed with a slowly varying average power request \cite{mocera2020analysis}. As such, the tractor and the powertrain work in almost constant operating points. Transport operations, on the contrary, are much more variable scenarios and so more challenging for the EMS. Simulation results revealed that the proposed MPC can achieve good saving performance without losing in speed tracking.

To summarize, the main contributions of this work are: 1) the experimental identification and validation of a control-oriented model of the tractor and its engine speed controller; 2) the formulation of an energy management strategy based on an MPC approach, which allows the explicit consideration the presence of the built-in ICE speed controller; 3) the performance validation -- in terms of fuel saving and speed tracking -- and the sensitivity analysis carried out on the experimentally-validated simulator, considering both transport and agricultural driving-cycles.

The remainder of the paper is organized as follows. In Section \ref{se_exp_mod}, the experimental setup is presented along with its model and identification procedure. In Section \ref{se_prob}, the energy management problem for the vehicle is introduced and then, in Section \ref{se_mpc}, formulated in the MPC framework, showing the cost function, the predictive model and some additional constraints. Finally, the simulation results are discussed in Section \ref{se_sim_res}.

\section{Experimental Setup and Simulation Environment}
\label{se_exp_mod}
\begin{figure}[t]
\centering
\includegraphics[width=8.5cm]{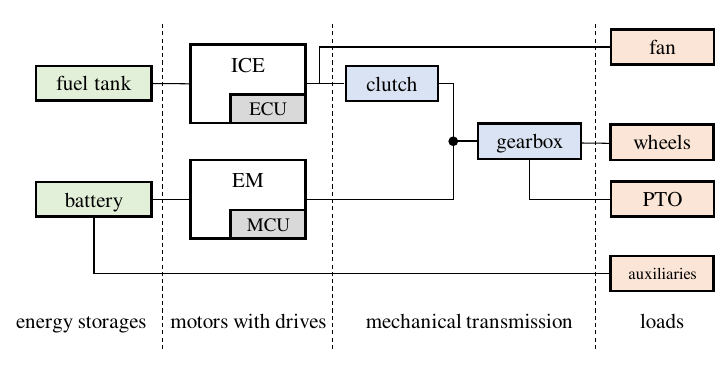}
\caption{Vehicle scheme.}
\label{fig_vehicle_scheme}
\end{figure}

The considered vehicle is an orchard/vineyard hybrid tractor prototype. Its traction system is composed of an electric motor (EM) connected with the battery through an inverter (MCU), that already includes a torque controller. The electric motor is mounted on the main shaft in parallel with the Diesel internal combustion engine equipped with its built-in control unit (ECU), oriented to the tracking of the driver speed reference. Moreover, the engine keeps a fan in rotation for cooling purposes. Between the main shaft and the wheel one, there is a manual transmission with twelve gearbox ratios. While the EM is always mechanically coupled with the transmission, the ICE can be decoupled by a controllable clutch. Typical of the tractors is the presence of the PTO, used to connect agricultural machines with the vehicle, and the presence of hydraulic auxiliaries, in this case, driven by a dedicated electric motor-pump system. The complete schematic representation of the vehicle is visualized in Fig. \ref{fig_vehicle_scheme}. 

In the following, a mathematical model of the vehicle is derived from experimental data in order to develop a simulator of the vehicle in Matlab/Simulink\footnote{MathWorks, Portola Valley, California, United States.} environment. Each element of the vehicle will be singularly characterized from data collected in different experiments, then the complete model will be validated on a transport maneuver. 
\subsection{Components modeling and identification}
\begin{figure}[h]
\centering
\includegraphics[scale=0.9]{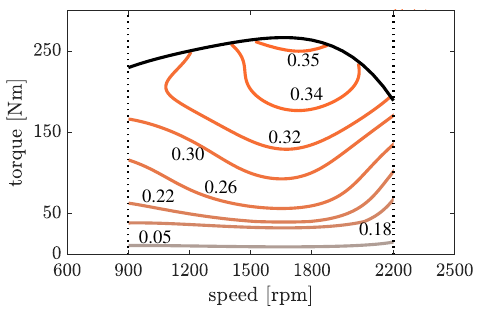}
\caption{ICE efficiency map. Engine speed limits are highlighted in dashed lines and maximum torque is in continuous line.}
\label{fig_ICE_eff}
\end{figure}

\textbf{Internal combustion engine.} The internal combustion engine is modeled as a static map \cite{rizzoni1999unified}, which represents the efficiency $\eta_\mathrm{ice}$ between the fuel power consumed $P_\mathrm{f}$ and the mechanical one $P_\mathrm{ice}$ in a certain operating point (rotational speed and torque provided):
\begin{equation}
\eta_\mathrm{ice}(T_\mathrm{ice},\Omega) = \frac{P_\mathrm{ice}}{P_\mathrm{f}} =  \frac{T_\mathrm{ice} \Omega}{\lambda_\mathrm{f}   \dot{m}_\mathrm{f}},
\label{eq_ICE_eff}
\end{equation}  
where $\Omega$ is the rotational speed of the main shaft and $T_\mathrm{ice}$ is the engine torque, that is the equivalent torque applied by the engine pistons reduced by the internal friction; therefore, when the pistons are not applying any torque, i.e., when the injected fuel is null, the net engine torque $T_\mathrm{ice}$ assumes negative values, because of the frictions. Then, $\dot{m}_\mathrm{f}$ is the fuel rate and $\lambda_\mathrm{f}$ is the lower heating value, that represents the energy density of the fuel provided during combustion. The map provided in Fig. \ref{fig_ICE_eff} is the result of a polynomial surface fitting on experimental efficiency data points, collected by keeping the engine in a fixed operating point ($\Omega,T_\mathrm{ice}$) with a dyno. \\

\begin{figure}[h]
\centering
\includegraphics[scale=0.9]{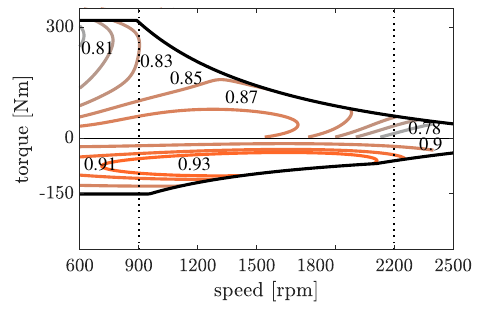}
\caption{EM efficiency. Torque limits are highlighted in continuous line, in dashed line the speed limits of the engine are reported to show the speed range possible in hybrid mode. The EM presents different limits in traction and recharge to satisfy the battery ones.}
\label{fig_EM_eff}
\end{figure}

\textbf{Electric motor.} Given the high bandwidth of the torque controller in the MCU, the electric motor can be modeled with an efficiency map as done for the engine. In this case, the efficiency is defined as \cite{rizzoni1999unified}: 
\begin{equation}
\eta_\mathrm{em}(T_\mathrm{em},\Omega) = {\left(  \frac{T_\mathrm{em} \Omega}{V_\mathrm{b}I_\mathrm{b,em}} \right)} ^ {\sign(T_\mathrm{em})}.
\label{eq_EM_eff_2}
\end{equation}
where $T_\mathrm{em}$ is the provided torque, $V_\mathrm{b}$ and $I_\mathrm{b,em}$ are the voltage and current at the battery side. It should be pointed out that the efficiency $\eta_\mathrm{em}$, as defined in \eqref{eq_EM_eff_2}, includes also the efficiency of the inverter, used to control the motor. Electric machines can also generate electrical power when applying negative braking torques, therefore the $\sign(T_\mathrm{em})$ appears in \eqref{eq_EM_eff_2}. Fig. \ref{fig_EM_eff} shows the EM efficiency; following the same procedure carried out for the engine identification, it is the result of a surface fitted on experimental data collected at the test bench. \\

\begin{figure}[h]
\centering
\includegraphics[scale=0.9]{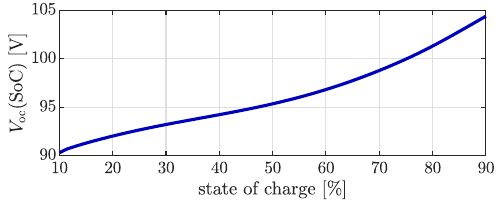}
\caption{SoC - Voltage battery characteristic.}
\label{fig_V_SoC}
\end{figure}

\textbf{Battery.} Battery is modeled according to its equivalent circuit model, that is an ideal voltage generator $V_\mathrm{oc}$ in series with a resistance $R_\mathrm{b}$:
\begin{equation}
V_\mathrm{b} = V_\mathrm{oc}(\text{SoC}) - R_\mathrm{b}I_\mathrm{b},
\label{eq_batt}
\end{equation}
where $I_\mathrm{b}$ is the sum of the current delivered to the motor $I_\mathrm{b,em}$ and the one required by the auxiliaries $I_\mathrm{b,aux}$.
Moreover, the open-circuit $V_\mathrm{oc}$ of Li-ion batteries varies around its nominal voltage $V_\mathrm{n}$ as a function of the state of charge (SoC). The  shape of this function depends on the Li-ion cells that the battery is composed of, Fig. \ref{fig_V_SoC} corresponds to the one on the vehicle.
SoC can be defined as the available capacity of the battery with respect to the nominal one $Q_\mathrm{b}$ at full charge, therefore the SoC dynamics is ruled by the following equation:
\begin{equation}
\frac{\mathrm{d}\text{SoC}}{\mathrm{d}t} = - \frac{I_\mathrm{b}}{Q_\mathrm{b}}.
\label{eq_soc}
\end{equation}

\begin{figure}[h]
\centering
\includegraphics[scale=0.9]{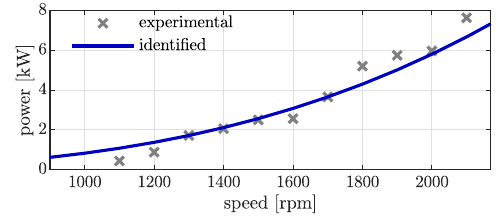}
\caption{ICE fan losses. Experimental data are compared with the identified model.}
\label{fig_ICE_fan}
\end{figure}

\textbf{Fan.} For cooling purposes, the engine keeps a fan in rotation, that is modeled as an additional load power $P_\mathrm{fan}$ as a function of the speed:
\begin{equation}
P_\mathrm{fan} = T_\mathrm{fan}\Omega = \left( A_\mathrm{fan}{\Omega}^2 + B_\mathrm{fan}\Omega +C_\mathrm{fan} \right)\Omega.
\label{eq_ICE_fan}
\end{equation}
where $A_\mathrm{fan}$, $B_\mathrm{fan}$ and $C_\mathrm{fan}$ are the model coefficients. Fig. \ref{fig_ICE_fan} shows the experimental data collected while keeping the ICE idling, without any load except for the fan and compared with the identified model.\\

\textbf{Transmission.}  Transmission in agricultural tractors is an oil-bath one, modeled through the power loss $P_\mathrm{gb}$ necessary to idle the transmission as a function of the transmission speed $\Omega_\mathrm{gb}=\Omega/\tau_\mathrm{gb}$ \cite{guzzella2013propulsion}, i.e., the speed $\Omega$ of the main shaft scaled by the gear ratio $\tau_\mathrm{gb}$. The gearbox is composed of twelve values divided into three sets: slow (S), medium (M) and fast (F). Moreover, due to the presence of oil, the losses decrease when the oil temperature $\Upsilon$ increases \cite{molari2008transmission}:
\begin{equation}
\def\arraystretch{1.2}
\begin{array}{ll}
P_\mathrm{gb} & = T_\mathrm{gb}(\Omega_\mathrm{gb},\Upsilon)\Omega_\mathrm{gb} = \\ 
&= \left( A_\mathrm{gb}(\Upsilon){\left(\frac{\Omega}{\tau\mathrm{gb}}\right)}^2 + B_\mathrm{gb}(\Upsilon){\left(\frac{\Omega}{\tau\mathrm{gb}}\right)} \right){\Big(\frac{\Omega}{\tau\mathrm{gb}}\Big)},
\end{array}
\label{eq_gb_loss}
\end{equation} 
where $A_\mathrm{gb}$ and $B_\mathrm{gb}$ are the model coefficients.

\begin{figure}[h]
\centering
\includegraphics[scale=0.9]{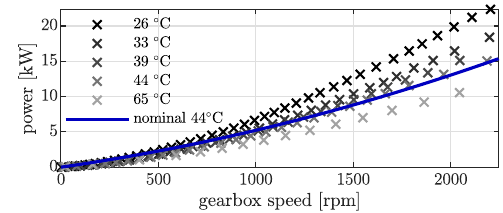}
\caption{Transmission losses. Experimental data collected at different temperatures are compared with the identified model at the nominal temperature of 44\textsuperscript{o}C.}
\label{fig_trans}
\end{figure}

The experimental data, collected at different temperatures, while the transmission is idling, are shown in Fig. \ref{fig_trans}. Then, the model is identified around the nominal temperature of 44\textsuperscript{o}C, i.e., the average operating temperature reached during the typical usage of the tractor.\\

\textbf{Longitudinal dynamics.} The longitudinal dynamics of the vehicle can be computed with a longitudinal power balance, that results in:
\begin{equation}
M\frac{\mathrm{d}v}{\mathrm{d}t}v = P_\mathrm{tr}-P_\mathrm{br}-P_\mathrm{cd}(v)-P_\mathrm{pto},
\label{eq_lon_model}
\end{equation} 
where $M$ is the vehicle mass and $v$ the longitudinal speed. $P_\mathrm{tr}$ is the equivalent traction power at the wheel, that is the sum of the two motors power reduced by the fan and transmission power losses. Then, $P_\mathrm{br}=F_\mathrm{br}v$ is the braking power requested by the driver. $P_\mathrm{pto}=T_\mathrm{pto}\Omega\tau_\mathrm{pto}$ is the power to be delivered to the loads connected through the PTO, where $T_\mathrm{pto}$ is the PTO torque and $\tau_\mathrm{pto}$ is the gear ratio between the PTO shaft and the engine one. Finally, $P_\mathrm{cd}(v)$ is the so-called coasting-down power to counteract when the vehicle is kept at constant speed $v$. It can be defined as the product between the longitudinal speed $v$ and the coasting-down force $F_\mathrm{cd}$, the force that must be exerted to drive the vehicle at a certain (constant) speed. It accounts for three components: the aerodynamic friction proportional to the speed squared, the viscous resistance proportional to the speed and a constant value as a function of the rolling resistance and the road slope:
\begin{equation}
	\begin{array}{ll}
		F_\mathrm{cd}(v) = \frac{1}{2}\rho A_\mathrm{x} C_\mathrm{x} v^2 + \beta v + Mg C_\mathrm{r}\cos\theta + Mg \sin\theta.
	\end{array}
	\label{eq_lon_model_cd0}
\end{equation}
Regarding the first term, $\rho$ is the air density, $A_\mathrm{x}$ the frontal surface area, and $C_\mathrm{x}$ the drag coefficient. $\beta$ is the viscous coefficient and $C_\mathrm{r}$ the rolling resistance coefficient. Finally, $M$ is the total vehicle mass, $g$ the gravity and $\theta$ the road slope, positive when up-hill. In place of using the values of these physical parameters, it is common to lump them into three coefficients --  $A = Mg C_\mathrm{r}, B = \beta$ and $C = \frac{1}{2}\rho A_\mathrm{x} C_\mathrm{x}$ -- which can be easily experimentally identified \cite{guzzella2013propulsion}:
\begin{equation}
	\begin{array}{ll}
		F_\mathrm{cd}(v) = C v^2 + B v + A\cos\theta + Mg \sin\theta. 
	\end{array}
	\label{eq_lon_model_cd}
\end{equation}

The longitudinal vehicle speed $v$ is related to the engine rotational speed according to:
\begin{equation}
v = \frac{\Omega}{\tau_\mathrm{gb}\tau_{0}}R_\mathrm{w},
\label{eq_v_omega}
\end{equation} 
where $\tau_0$ is the fixed ratio between the rotational speed of the gearbox and $R_\mathrm{w}$ is the wheel radius.

\begin{figure}[h]
\centering
\includegraphics[scale=0.9]{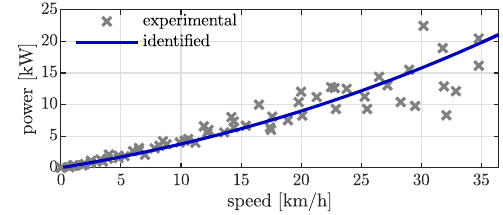}
\caption{Coasting-down losses. Experimental data are compared with the identified model.}
\label{fig_cd_loss}
\end{figure}

The coefficients of the coasting-down force in \eqref{eq_lon_model_cd} can be experimentally found letting decelerating the vehicle on its own on a flat road, with the clutch open and the PTO disconnected. Hence, \eqref{eq_lon_model} becomes:
\begin{equation}
M\frac{\mathrm{d}v}{\mathrm{d}t}v= -P_\mathrm{gb}-P_\mathrm{cd}(v),
\label{eq_lon_model_cd_only}
\end{equation} 
and therefore the model can be fitted on the experimental data as shown in Fig. \ref{fig_cd_loss}. During the identification tests, speed and acceleration have been measured with an additional GPS/IMU system, the mass is considered known and the gearbox losses known from Fig. \ref{fig_trans}. Given that the fan is placed before the clutch, the fan power losses do not appear in \eqref{eq_lon_model_cd_only}.  \\

\begin{figure}[h]
\centering
\includegraphics[scale = 0.8]{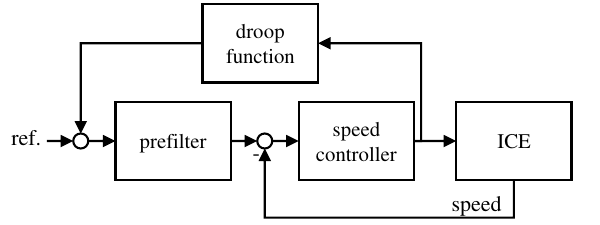}
\caption{ICE control scheme model.}
\label{fig_ICE_control_scheme}
\end{figure}

\textbf{Engine controller.} The last element to be presented is the engine speed controller. In fact, with respect to traditional vehicles, where the driver gives a torque command to the engine through the throttle pedal, in agricultural ones, the driver sends a rotational speed reference for the engine itself \cite{lee2019engine,mocera2021model}. The control scheme of the engine is shown in the scheme in Fig. \ref{fig_ICE_control_scheme}, where two additional elements are visible: the 1.5 Hz reference prefilter and the droop function, that modifies the speed reference requested by the driver as a function of the current torque applied by the engine \cite{garvey1983droop}. In particular, the reference increases when the load reduces, in order to have a smoother control action in load transitions and moreover the driver is able to experience a load variation even if the engine is speed controlled. In our application, the droop function has been disabled, so that the engine does not present any steady-state error with respect to the driver request.

Deriving a detailed model of the speed controller is a hard task, in fact control of diesel engines is very complex, due to the multiple objectives (e.g., injection, air flow, temperature, emission), in addition to the main one, that keeps the engine in the desired operating point \cite{guzella1998diesel,guzzella2010ice}. For the purposes of this work, the controller is modeled in a control-oriented way. Hence, it is simplified as a PI (proportional-integral) engine speed controller, whose equations are:
\begin{equation}
\left\{\begin{array}{l}
T_\mathrm{ice} = K_\mathrm{p}(\Omega_\mathrm{ref}-\Omega)+x_\mathrm{ice}\\
\dot{x}_\mathrm{ice} = K_\mathrm{i}(\Omega_\mathrm{ref}-\Omega)
\end{array}\right.,
\label{eq_ICE_controller}
\end{equation}  
where $\Omega_\mathrm{ref}$ is the filtered reference from the pedal, $K_\mathrm{p}$ and $K_\mathrm{i}$ the PI gains and $x_\mathrm{ice}$ the controller status associated to the integral action. The experimental identification of the control gains is discussed after the definition of the complete model equations.

\subsection{Complete vehicle modeling and identification}
The complete model, derived by merging the models of each vehicle component \eqref{eq_ICE_eff}--\eqref{eq_ICE_controller}, results in the following set of equations:
\begin{equation}
\small
\def\arraystretch{1.1}
\left\{\begin{array}{lcl}

\dot{v} &=& \frac{1}{M R_\mathrm{w}}\Big[ \big((T_\mathrm{ice}-T_\mathrm{fan})\xi + T_\mathrm{em}-T_\mathrm{pto}\tau_\mathrm{pto}\big)\tau_\mathrm{gb}\tau_0 +\\
&&- (F_\mathrm{cd}-F_\mathrm{br})R_\mathrm{w} - T_\mathrm{gb}\tau_0 \Big]\\

\dot{m}_\mathrm{f} &=& \frac{1}{\lambda_\mathrm{f}}\left(\frac{T_\mathrm{ice}\Omega}{\eta_\mathrm{ice}}\right)\xi\\
T_\mathrm{ice} &=& K_\mathrm{p}(\Omega_\mathrm{ref}-\Omega)+x_\mathrm{ice}\\
\dot{x}_\mathrm{ice} &=& K_\mathrm{i}(\Omega_\mathrm{ref}-\Omega)\\

\text{S}\dot{\text{o}}\text{C} &=& - \frac{I_\mathrm{b}}{Q_\mathrm{b}}\\
V_\mathrm{b} &=& V_\mathrm{oc} - R_\mathrm{b}I_\mathrm{b}\\
I_\mathrm{b} &=& \frac{T_\mathrm{em}\Omega}{V_\mathrm{b}\eta_\mathrm{em}} + I_\mathrm{b,aux}\\
\end{array}\right.
\label{eq_model}
\end{equation}  
where $\xi$ is a boolean variable representing the status of the clutch:
\begin{equation}
\xi = \left\{\begin{array}{ll}
1 & $when closed$\\
0 & $when open$
\end{array}\right.
\label{eq_clutch}
\end{equation}
Considering an energy-oriented modeling framework, the role of the clutch is simplified into a boolean status, in order to model the possibility of engine off, with $\xi= $0, driving the vehicle in full-electric mode.

This model has four continuous states: speed, fuel consumption, SoC and the integral action of the speed controller; and four inputs: the engine speed reference, the electric motor torque, the clutch status and the gearbox ratio. Braking force, PTO torque and the current required by the auxiliaries acts as disturbances. On this tractor, the braking force and auxiliaries' current are measured, while the PTO depends on the connected machinery.\\
 
\begin{figure}[h]
\centering
\includegraphics[scale = 0.9]{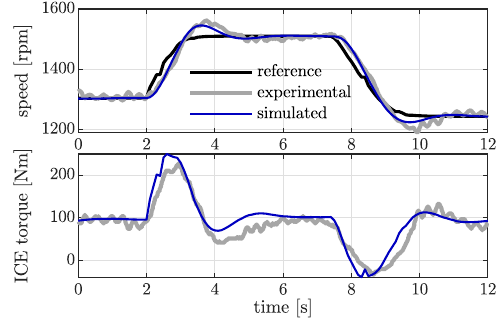}
\caption{ICE controller model validation. Data have been collected during an ICE-only test.}
\label{fig_val_ice}
\end{figure}

Once the complete model is derived, the controller gains can be identified, performing multiple speed reference steps and minimizing the mismatch between the experimentally measured engine speed ($\Omega^\mathrm{meas}$) and the one obtained by running the model \eqref{eq_model} ($\Omega^\mathrm{sim}$):
\begin{equation}
(K_\mathrm{p},K_\mathrm{i}) = \mathrm{argmin}  \sum_{i} \left( \Omega^\mathrm{meas}(i) - \Omega^\mathrm{sim}(i)\right)^2.
\end{equation}
The comparison of the measured and simulated data with the optimal controller gains found is shown in Fig. \ref{fig_val_ice}. This figure highlights the capability of the PI model to match the simulated speed with the measured one in terms of the rise-time, overshoot and steady-state error. Also the good match between the experimental and simulated control variable is provided as validation.

\subsection{Simulation environment validation}
\begin{figure*}[t]
\centering
\includegraphics[scale=0.9]{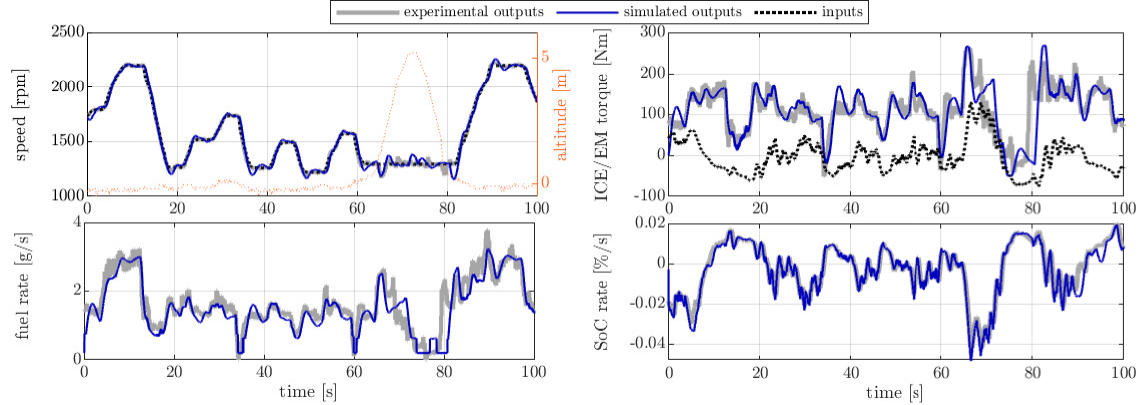}
\caption{Complete model validation. Experimental data (gray) are compared with the simulator outputs (blue) driven by the same experimental inputs (black dotted). The orange dotted line in the first plot represent the altitude.}
\label{fig_val_tot}
\rule[1ex]{\textwidth}{0.1pt}
\end{figure*} 
The model in \eqref{eq_model} is implemented in Matlab/Simulink as simulation environment for the energy management strategies. The complete validation has been carried out by comparing experimental data and simulated ones when fed by the same inputs. The inputs of the test are the speed reference, imposed by the driver; the electric motor torque, required by the vehicle control unit which was operating following a torque-assist rationale; the altitude. During this test, the clutch is always closed and the gear ratio is constant over the entire maneuver. Considering the transport maneuver, no additional loads were present in the test. The model validation, see Fig. \ref{fig_val_tot}, is assessed by the good match between the simulated and measured outputs, i.e., the engine torque regulated by the speed controller, the SoC rate and the fuel rate. Among the several, these variables have been selected since they are (or are tightly related) to the vehicle model states. In particular, the matching speed and engine torque validate the controller gains, engine torque also validates the models of the vehicle dynamics and loads (fan, transmission), while fuel and SoC rates validate the ICE and EM efficiency maps, respectively. The main model parameters are shown in Tab. \ref{tab_param}.
\begin{table}[h]
\centering
\caption{main vehicle parameters}
\label{tab_param}
\begin{tabular}{ccc}
\hline
\textbf{parameter} & \textbf{value} & \textbf{unit} \\
\hline
\hline
$M$ & 3830 & kg \\
$R_\mathrm{w}$ & 65 & cm \\
$\tau_0$ & 22.8 & $-$ \\
$K_\mathrm{p}$ & 12 & Nms/rad \\
$K_\mathrm{i}$ & 3.33 & Nm/rad \\
$\lambda_\mathrm{f}$ & 42.68 & MJ/kg \\
$Q_\mathrm{b}$ & 14 & kWh \\
\hline
\end{tabular}
\end{table}

\section{Problem Formulation}
\label{se_prob}
The energy management problem for this vehicle is addressed in order to minimize fuel consumption, improving the vehicle efficiency, while tracking the speed reference requested by the driver. Moreover, considering that the vehicle is non plug-in, the energy management should be able to guarantee the battery charge-sustaining.
In traditional parallel hybrid vehicles, the combined energy management and cruise control is typically addressed by hierarchical control schemes, e.g., \cite{kural2015emsacc,he2020emsacc,liu2022emsacc}, where an external controller computes the total torque necessary to track the desired speed and an internal loop, the proper EMS, splits the requested torque between the engine and the electric motor in order to minimize consumption.

\begin{figure}[h]
\centering
\includegraphics[scale=0.85]{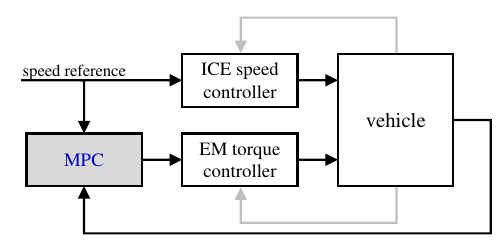}
\caption{Vehicle control scheme model.}
\label{fig_vehicle_control_scheme}
\end{figure}

An important difference between tractors and traditional vehicles is the speed-controlled internal combustion engine. With such a configuration, the delivered ICE torque cannot be directly manipulated by the EMS, unless the ICE control architecture is completely revised. Sometimes, for industrial constraints, this is not even possible. In our work, we address this challenge and propose an EMS for a parallel hybrid vehicle that manipulates only the electric motor torque and does not (at least directly) interfere with the ICE control architecture, see Fig. \ref{fig_vehicle_control_scheme}. The high-level control goal, i.e., the speed reference tracking, is still ultimately fulfilled by the engine, whereas the primary goal of the electric motor becomes the efficiency improvement.

The proposed add-on solution is an MPC-based energy management strategy. Indeed, MPC seems a promising framework for two reasons: 1) the multi-objective nature of the problem can be addressed by introducing both fuel consumption and speed tracking error minimization in the cost function; 2) the presence of the built-in speed controller can be included in the vehicle model present in the MPC.

\section{MPC-based Energy Management Strategy}
\label{se_mpc}
To describe the MPC, the cost function to minimize, the predictive model and the eventual additional constraints are introduced.
Then, the solver used is presented and the tuning of the prediction horizon length and the weights of the cost function contributions is discussed.\\

\textbf{Cost function.} The proposed cost function is the sum of four elements, the first two are associated with the energy consumption minimization goal, the third one is oriented to the speed tracking, while the last one is an auxiliary term used to have a smooth control variable:
\begin{equation}
\begin{array}{ccl}
J & = & c_\mathrm{fuel}m_\mathrm{f}(N) + c_\mathrm{soc}\left[\text{SoC}(N)-\text{SoC}(1)\right] +\\
  &   & c_\mathrm{track}\sum_{k=1}^{N}{\left(\Omega(k)-\Omega_\mathrm{ref}(k)\right)^2} +\\
  &   & c_\mathrm{control}\sum_{k=1}^{N-1}{\left(T_\mathrm{em}(k+1)-T_\mathrm{em}(k)\right)^2}
\end{array}.
\label{eq_cost_fun}
\end{equation}
$N$ is the prediction horizon length and $c_{*}$ are the weights on each contribution. The first two weights are chosen in a physical way, in fact they are used to link fuel and SoC consumption to energy consumption, therefore the following equations hold:
\begin{equation}
\begin{array}{ccccc}
c_\mathrm{fuel} = \lambda_\mathrm{f} & \left[\frac{\text{J}}{\text{kg}}\right] & \text{ and } & c_\mathrm{soc}  = \frac{V_\mathrm{n}Q_\mathrm{b}}{100} & \left[\frac{\mathrm{J}}{\%}\right]
\end{array},
\end{equation}
while other parameters will be tuned according to a sensitivity analysis presented below.\\

\textbf{Predictive model.} The predictive model is a simplified and discretized version of \eqref{eq_model}. The first simplification consists of replacing model equation \eqref{eq_batt} with the nominal battery voltage. This simplification is reasonable for charge-sustaining vehicles, indeed the battery SoC remains around the nominal value \cite{onori2016hybrid}. The second one makes the EM torque the only optimization variable, linking the clutch status with the engine torque: the clutch opens, activating the full-electric mode, when the engine torque is close to zero ($T_\mathrm{ice} < T_\mathrm{th} \leftrightarrow \xi=0$). Then, the model is written in discrete time according to the forward Euler method with sample time $\Delta_\mathrm{s} = $ 0.05 s, chosen to have 10 samples in system rise time.

\begin{figure*}
\begin{equation}\tag{OP1}
\def\arraystretch{1.2}
\begin{array}{cll}
\min\limits_{T_\mathrm{em},X} & c_\mathrm{fuel}m_\mathrm{f}(N) + c_\mathrm{soc}\left[\text{SoC}(N)-\text{SoC}(1)\right] +\\
  &  c_\mathrm{track}\sum_{k=1}^{N}{\left(\Omega(k)-\Omega_\mathrm{ref}(k)\right)^2} +\\
  &  c_\mathrm{control}\sum_{k=1}^{N-1}{\left(T_\mathrm{em}(k+1)-T_\mathrm{em}(k)\right)^2}\\
$s.t$. & $model constraints$ & \\
 & X(k+1) = X(k)+\Delta_\mathrm{s} F(X(k),T_\mathrm{em}(k),\tau_\mathrm{gb}(k),\Omega_\mathrm{ref}(k),T_\mathrm{pto}(k),I_\mathrm{b,aux}(k))& k=1,...,N-1 \\
 & X(1) = X^\mathrm{meas} & \\
 & \tau_\mathrm{gb}(k) = \tau_\mathrm{gb}^\mathrm{meas} & k=1,...,N  \\
 & \Omega_\mathrm{ref}(k+1) = \Omega_\mathrm{ref}(k)+\Delta_\mathrm{s}\dot{\Omega}_\mathrm{ref}^\mathrm{meas} & k=1,...,N-1\\
 & $hard constraints$ & \\
 & \text{SoC}^\mathrm{min} \leq \text{SoC}(k) \leq \text{SoC}^\mathrm{max} & k=1,...,N  \\
 & T_\mathrm{em}^\mathrm{min}(\Omega) \leq T_\mathrm{em}(k) \leq T_\mathrm{em}^\mathrm{max}(\Omega) & k=1,...,N \\
 & \frac{I_\mathrm{b}^\mathrm{min}}{Q_\mathrm{b}} \leq -\frac{\text{SoC}(k+1)-\text{SoC}(k)}{\Delta_\mathrm{s}} \leq \frac{I_\mathrm{b}^\mathrm{max}}{Q_\mathrm{b}} & k=1,...,N-1  \\
\end{array}
\label{eq_op}
\end{equation}
\rule[1ex]{\textwidth}{0.1pt}
\end{figure*}

In conclusion, the predictive model is a discrete time nonlinear system that can be written in compact form as:
\begin{equation}
\begin{array}{ll}
X(k+1) & = X(k)+\Delta_\mathrm{s}F(X(k),T_\mathrm{em}(k),\tau_\mathrm{gb}(k),\Omega_\mathrm{ref}(k)) \\
X(1) & = X^\mathrm{meas}
\end{array},
\label{eq_pred_mod}
\end{equation}
where $X = [\Omega,m_\mathrm{f},\text{SoC},x_\mathrm{ice}]$ is the vector of the systems states, initialized with the last available measurement (superscript $\mathrm{meas}$), and $F$ is the state dynamics function. It is visible that the model is a function of two unknown inputs: the speed reference and the gearbox ratio. The first one is predicted keeping its derivative constant over the horizon:
\begin{equation}
\Omega_\mathrm{ref}(k+1) = \Omega_\mathrm{ref}(k)+\Delta_\mathrm{s}\dot{\Omega}_\mathrm{ref}^\mathrm{meas},
\end{equation} 
while the gearbox ratio is kept constant:
\begin{equation}
\tau_\mathrm{gb}(k) = \tau_\mathrm{gb}^\mathrm{meas}.
\end{equation} 
This model can be introduced in the optimization model as a set of $N$ equality constraints and using $X(k)$ as a set of auxiliary optimization variables, according to the so-called multiple shooting method \cite{kiehl1994multiple}.  \\

\textbf{Constraints.} The problem is subject to three hard constraints:
\begin{itemize}
\item[1)] SoC is limited to stay between two levels in order to prevent the complete discharge and charge of the battery: 
\begin{equation}
\text{SoC}^\mathrm{min} \leq \text{SoC}(k) \leq \text{SoC}^\mathrm{max};
\end{equation}
\item[2)] EM torque is limited at each speed by its maximum value in charge and recharge: 
\begin{equation}
T_\mathrm{em}^\mathrm{min}(\Omega) \leq T_\mathrm{em}(k) \leq T_\mathrm{em}^\mathrm{max}(\Omega);
\end{equation}
\item[3)] battery current limits are suggested in real-time by the battery management system in order to preserve the battery  health. To avoid including the nonlinearities of the current expression in \eqref{eq_model}, the constraint can be written equivalently on the SoC rate, thanks to \eqref{eq_soc}:
\begin{equation}
\frac{I_\mathrm{b}^\mathrm{min}}{Q_\mathrm{b}} \leq -\frac{\text{SoC}(k+1)-\text{SoC}(k)}{\Delta_\mathrm{s}} \leq \frac{I_\mathrm{b}^\mathrm{max}}{Q_\mathrm{b}}.
\end{equation}
\end{itemize}

\vspace{10pt}\textbf{Solver design.} MPC formulates an optimization problem (OP) that can be summed up as in \eqref{eq_op}. This problem is solved with the support of CasADi \cite{andersson2019casdi}, which interprets the problem as a symbolic one and turns it into a numeric one at each iteration, given the measurements from the system. The OP is nonlinear, therefore a suitable solver is interior point (IPopt) \cite{potra2000ipopt}. It needs an initial guess as starting point to compute the optimal value, that is chosen to be coincident with the optimal time-shifted trajectory computed in the previous iteration \cite{gros2020initialpoint}, except for the very first iteration that it is set to zero.\\

\begin{figure}[h]
\centering
\includegraphics[scale=0.9]{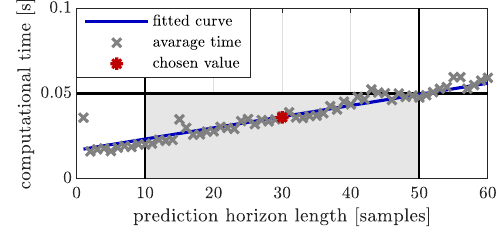}
\caption{Sensitivity of the computation time $\Delta_c$ w.r.t. to the horizon length $N$. The chosen value is highlighted, showing the satisfaction of the constraints.}
\label{fig_N}
\end{figure}

\textbf{Prediction horizon.} Prediction horizon $N$ is chosen through a sensitivity analysis with respect to the computational time $\Delta_\mathrm{c}$ needed to solve the OP in \eqref{eq_op}. Fig. \ref{fig_N} shows the average computational time starting from 50 random starting conditions (measured variables) of the OP. 
The chosen value is $N=$ 30, that corresponds to a prediction horizon of 1.5 s, in order to satisfy the following constraints: 1)
$N>$ 10, to include the transients of controller in the predictive model; 2) $\Delta_\mathrm{c}<$ 0.05 s, to guarantee the computation of the control variable at each sample time of the discrete time model, and apply the first optimal value on the system. \\

\begin{figure}[t]
\centering
\includegraphics[scale=0.9]{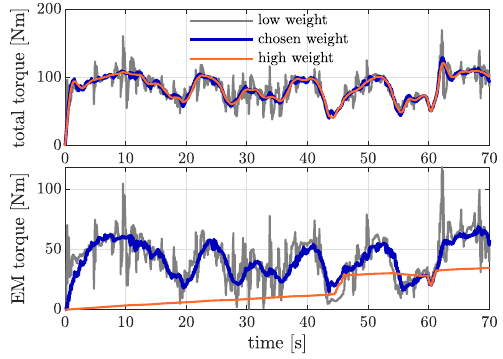}
\caption{Sensitivity w.r.t. the control weight $c_{control}$, in simulation environment.}
\label{fig_control_sens}
\end{figure}

\begin{figure}[!h]
\centering
\includegraphics[scale=0.9]{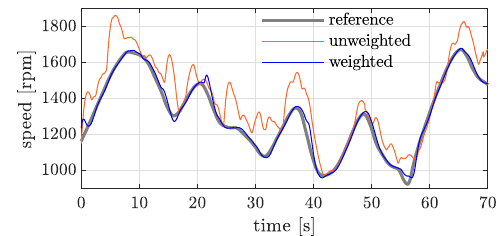}
\caption{Sensitivity w.r.t. the control weight $c_{track}$, in simulation environment.}
\label{fig_track_sens}
\end{figure}

\textbf{Cost function weights.} Once the framework to solve the MPC problem is defined, it is possible to perform sensitivity analyses to tune the remaining parameters: the weights on the tracking error $c_\mathrm{track}$ and on the control variable $c_\mathrm{control}$. $c_\mathrm{control}$ is chosen in order to manage the trade-off, visible in Fig. \ref{fig_control_sens}; recalling that the total torque to follow the reference speed is adjusted by the ICE speed controller, it is possible to conclude that: when the weight is too low, the EM and the total torque are too rough; while when it is too high, the total torque is smooth, but the electric motor is not as fast as the engine controller dynamics and therefore it provides a slow response. Finally, a good tuning of the parameter $c_\mathrm{control}$ provides that both EM and total torque are smooth, without cutting out the dominant frequencies of the EM one.
The other weight, $c_\mathrm{track}$ is necessary to guarantee a good tracking performance, given that the electric motor torque acts as a disturbance for the engine speed controller. Moreover, it plays a very important role in full-electric mode, when MPC becomes the only controller responsible for tracking. Fig. \ref{fig_track_sens} shows how the tracking performance with the chosen weight is very close to the response of the ICE controller when the electric motor is turned off, while a significant error is present when the tracking weight is too low. In this analysis, the tuning of the tracking weight is based on the tracking effect only, while in Section \ref{se_sim_res}, the trade-off between tracking and energy saving is discussed in detail. \\

\begin{figure}[h]
\centering
\includegraphics[scale=0.9]{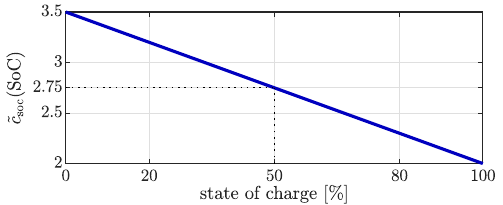}
\caption{Penalty function on the battery consumption as a function of SoC.}
\label{fig_k}
\end{figure}

\textbf{Charge-sustaining.} Up to now, the charge-sustaining constraint has been expressed as a hard constraint on the minimum and maximum value of the battery state of charge. Nevertheless, this solution is not sufficient, in fact in a charge-sustaining scenario the behavior of the energy management strategy should be SoC dependent, preferring a battery discharge when the SoC is too high and vice-versa \cite{paganelli2002ecms,onori2010aecms}. A possible solution to manage this problem is to introduce a penalty function on the battery consumption \cite{kleinmaier2002penalty}, such that the battery usage becomes more expensive when the SoC decreases. Therefore, the $c_\mathrm{soc}$ weight in the cost function \eqref{eq_cost_fun} is multiplied by a penalty function $\tilde{c}_\mathrm{soc}(\text{SoC})$, shown in Fig. \ref{fig_k}. Moreover, it must be noticed that its average value corresponds to the average ratio between  $\eta_\mathrm{em}$ and $\eta_\mathrm{ice}$, in order to consider that each battery consumption requires past or future fuel consumption to maintain the battery state of charge \cite{onori2016hybrid,radrizzani2022emp}, when dealing with non plug-in vehicles. 

\section{Simulation Results}
\label{se_sim_res}
\begin{figure}[h]
\centering
\includegraphics[scale=0.9]{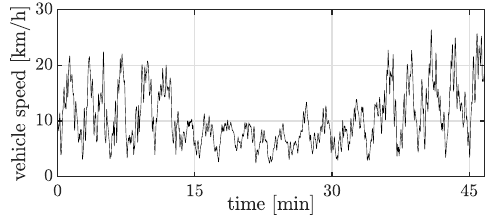}
\caption{Vehicle speed reference used in simulations.}
\label{fig_ref}
\end{figure}

The performance of the MPC controller is evaluated in the simulation environment developed in Section \ref{se_exp_mod}. As discussed in the Introduction, we mainly considered the transport scenario. In particular, the considered driving-cycle has been created using real measurements where different speeds, gears and driving patterns were recorded. In order to obtain a long driving-cycle (about 45 min), several surrogate profiles have been connected, obtained from the \textit{phase randomization process} of different shorter experimental driving-cycle. The phase randomization creates a surrogate signal by transforming the original one into the frequency domain and randomizing its phases, before reconverting into time domain \cite{theiler1992random}. The resulting speed reference is visible in Fig. \ref{fig_ref}. By applying this strategy, the reference speed has a power spectrum magnitude equal to measured shorter driving-cycles, avoiding to have unrepresentative accelerations and decelerations. The associated gear ratio is computed as a function of the reference vehicle speed.

Due to the multi-objective nature of the MPC, the trade-off between fuel saving and speed tracking is discussed through a sensitivity analysis. Moreover, the robustness with respect to the engine speed controller uncertainty is evaluated, in order to understand the effect of a possible mismatch with the model. Finally, we discuss the application of the proposed solution to agricultural applications.\\

\textbf{Performance evaluation.} The indexes used for performance evaluation are related to the two primary objectives of the MPC:  the fuel saving with respect to the ICE-only case (in this case, the vehicle mass is reduced by 200 kg, to consider the absence of the electric motor and the battery) and the speed tracking, quantified by the root mean square error (RMSE), normalized with respect to the speed reference. Results are also compared with the optimal global solution, considering the whole driving-cycle known. The global optimum can be computed via Pontryagin's minimum principle (PMP). Indeed, when hybrid vehicles operate in a charge-sustaining scenario, the battery voltage can be assumed almost constant, and if the PMP solution exists, it returns the global optimum, as discussed in \cite{serrao2009ecmspmp} and \cite{kim2012pmp}.

\begin{figure}[h]
\centering
\includegraphics[scale=0.9]{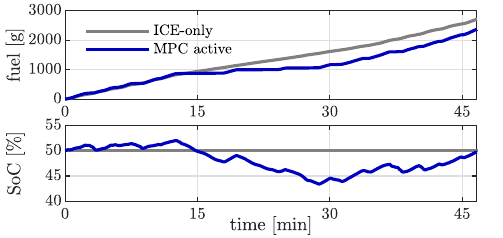}
\caption{Simulation results: fuel consumption and SoC behavior comparison between the MPC-based solution and the traditional ICE-only vehicle.}
\label{fig_res}
\end{figure}

Fig. \ref{fig_res} shows that the MPC-based solution makes the SoC return at the end of the driving-cycle close to its initial value with a lower fuel consumption than the ICE-only solution. To compare results with the global optimum, PMP is computed so to have the same initial SoC at the end of the driving-cycle. Results are summarized in Table \ref{tab_perf}: the MPC-based EMS produces a fuel saving, with respect to the traditional ICE-only vehicle, of 12.6 \% on the considered driving-cycle, while the global optimum reaches a level of 17.3 \%. 

\begin{table}[h]
\footnotesize
\centering
\caption{simulation results \\ comparison between the MPC-based solution, the traditional ICE-only vehicle and the optimal consumption considering the driving-cycle in Fig. \ref{fig_ref}}
\begin{tabular}{ccccc}
\hline
 & \textbf{tracking RMSE} & \textbf{fuel consumption} & \textbf{fuel saving} \\
 \hline
 \hline
\textbf{ICE-only} & 0.73 \% & 2704 g & -- \\
\textbf{MPC active} & 2.00 \% & 2364 g &  12.6 \% \\
\textbf{global optimum} & 0.76 \% & 2236 g &  17.3 \% \\
\hline
\end{tabular}
\centering
\label{tab_perf}
\end{table}

\begin{figure}[!h]
\centering
\includegraphics[scale=0.9]{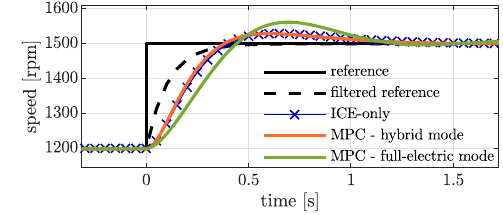}
\caption{The step response of the built-in engine controller is compared with the response when the MPC is active. In one case, it operates in hybrid mode and in one case in full-electric mode.}
\label{fig_step_comp}
\end{figure}

Analyzing results in terms of tracking error, it is possible to appreciate that all solutions do not deteriorate the tracking performance, quantified with a low tracking RMSE. Indeed, MPC generates an absolute error of 44 rpm at the top speed of 2200 rpm.
Despite the low value (2.0 \%) of the speed tracking RMSE in case of the MPC-based solution, it is higher than the original ICE-only solution and so it is interesting to analyze the motivations behind that. The main cause can be attributed to the full-electric mode, where the ICE speed controller is turned off and the MPC becomes entirely responsible for the speed tracking. Fig. \ref{fig_step_comp} compares the step response of the built-in engine controller with the response when the MPC is active. In one case, MPC operates in hybrid mode (recharging the battery) in parallel with the engine controller, in the other case in full-electric mode. All responses present the same settling-time, but in full-electric there is a slower rise-time and higher peak, because of the maximum torque reduction (see Fig. \ref{fig_EM_eff} and \ref{fig_ICE_eff}) and the slower sampling time of the control variable. Indeed, the built-in engine speed controller operates at 100 Hz, while the MPC sampling time is 20 Hz. Nevertheless, the performance can be considered satisfying, as shown by Fig. \ref{fig_step_comp} on a step response and evaluated by the speed error RMSE on the complete driving-cycle. Moreover, Fig. \ref{fig_step_comp} shows how the steady-state error in full-electric mode remains negligible, without an explicit MPC integral action, thanks to the presence of the integral action in the predictive model equations.

\begin{figure}[t]
\centering
\includegraphics[scale=0.9]{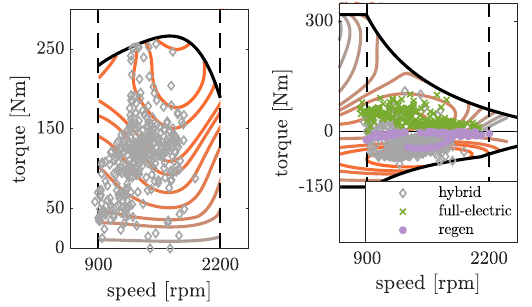}
\caption{Simulation results: engine (left) and electric motor (right) torque projected on the respective efficiency maps.}
\label{fig_points}
\end{figure}

Another interesting point to be discussed is how the MPC behaves during the driving-cycle, i.e., how the electric motor torque moves the engine one in order to save fuel. Figure \ref{fig_points} shows the operating points, highlighting a main pattern during the whole driving-cycle: 1) the full-electric mode is preferred at low loads, where the engine efficiency is very low; 2) when the hybrid mode is active, the EM torque is principally negative, recharging the battery, to move up the ICE operating point in more efficient zones. Finally, it is also visible that in the transport scenario, it is possible to recuperate energy, thanks to the regenerative braking of the electric motor. Indeed, given that the speed tracking is an objective of the MPC, the EMS automatically asks for negative torques to slow down the vehicle, when the speed is higher than the reference requested by the driver.

\begin{figure}[!h]
\centering
\includegraphics[scale=0.9]{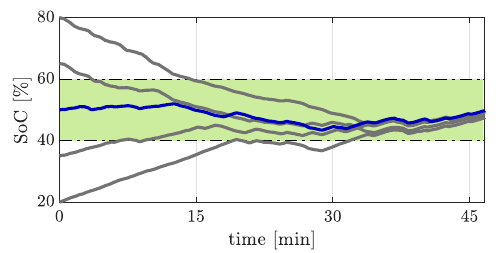}
\caption{SoC evolutions for different initial value converge to the desired SoC region, maintaining the battery SoC around the 50 \%.}
\label{fig_socs}
\end{figure}

The last objective of the MPC is the charge-sustaining capability: Fig. \ref{fig_socs} shows how the battery SoC converges to a similar behavior even if the initial values is different, thanks to the fact that at low SoC values the recharge is defined (according to the penalty function in Fig. \ref{fig_k}) as more convenient and vice-versa.\\

\begin{figure}[!h]
\centering
\includegraphics[scale=0.9]{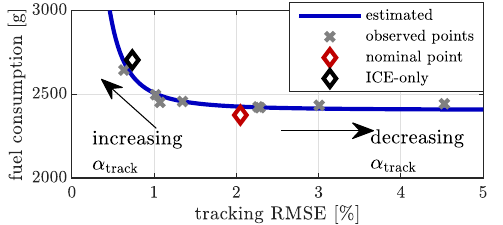}
\caption{Pareto curve fitted on experience data varying $\alpha_\mathrm{track}$. The chosen value at $\alpha_\mathrm{track}$=1 is highlighted in red.}
\label{fig_sens_alpha_track}
\end{figure}

\textbf{Saving/tracking trade-off.} The tuning of the tracking weight $c_\mathrm{track}$ has been validated through the additional sensitivity analysis in Fig. \ref{fig_track_sens}, with deeper analysis of its effects on the fuel saving. Considering again the driving-cycle in Fig. \ref{fig_ref}, the MPC tracking weight $c_\mathrm{track}$ is replaced by $ c_\mathrm{track}\alpha_\mathrm{track}$ in order to perform a sensitivity analysis increasing and reducing the chosen tracking weight. In Fig. \ref{fig_sens_alpha_track}, the Pareto curve resulting from different simulation experiments is reported, showing that a very high weight on the tracking disrupts the energy saving performance; however, with lower values, the energy performance is significantly higher with a negligible increase of tracking error. This figure shows also the very good position on the Pareto curve considering the value of $c_\mathrm{track}$ chosen in Fig. \ref{fig_track_sens}.\\ 

\begin{figure}[!h]
\centering
\includegraphics[scale=0.9]{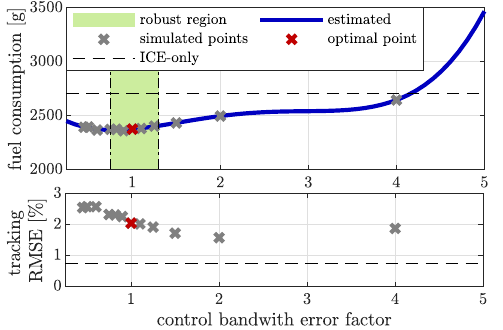}
\caption{Sensitivity of fuel consumption and tracking performance with respect to $\alpha_{\omega}$: fuel consumption increases out of the robust region when $\alpha_{\omega}$ differs from the nominal value, while tracking error is always kept in a small region.}
\label{fig_sens_alpha_omega}
\end{figure}

\textbf{Robustness to speed bandwidth model error.} The peculiarity of the proposed MPC is the interaction with the built-in engine speed controller, that is also included in the predictive model. In the previous simulation, the control model coincides with the real controller active on the plant; now, we are aiming at analyzing the performance loss due to a mismatch between the bandwidth of the controller in the model and the real one.
In particular, to change the control bandwidth, regulator gains are scaled in the model with a multiplicative factor, $\left(K_\mathrm{p},K_\mathrm{i}\right)$ is replaced by $\left(K_\mathrm{p}\alpha_{\omega},K_\mathrm{i}\alpha_{\omega}\right)$. The outcome of this analysis, in terms of fuel consumption and tracking RMSE, is both shown in Fig. \ref{fig_sens_alpha_omega}. First of all, it is visible that the tracking RMSE is almost invariant with respect to $\alpha_{\omega}$, due to the fact that the real control bandwidth does not change. Considering the fuel saving performance, it is possible to highlight a robust region, in fact, despite a model error, the performance is kept to the nominal level. On the other hand, when the model error increases the fuel consumption is visibly higher, due to the mismatch between the actual engine torque and the one present in the model. \\

\textbf{Agricultural applications.} The last analysis consists of the evaluation of the proposed MPC in agricultural operations. We recall that agricultural machinery occurs around a constant operating point \cite{mocera2020analysis}, which depends on the considered operation itself. The aim of this analysis is the validation the use of the MPC-based solution in such a scenario. Therefore, we selected a single driving-cycle, derived from experimental data. In particular, in the considered scythe operation, the speed reference was equal to 1200 rpm and the average load torque was 100 Nm. During this simulation, we assumed that the PTO torque was provided by a sensor, like the one shown in \cite{golinelli2019ptosensor}. However, if such a sensor is not available, load torque estimators can be employed \cite{kim2013loadTE}. 

We compared the fuel saving performance of the MPC with respect to the optimal global one, when the battery SoC returns to the initial value. In fact, thanks to the constant pattern of the requested speed and average load torque and the charge-sustaining scenario, the EMS will have a repetitive behavior, as well. Therefore, the simulation of long agricultural driving-cycles is not necessary to evaluate the fuel saving performance. 

Results showed that, in the considered operating point, the MPC can save the 5.40 \% of fuel with respect to the ICE-only tractor, which is close to the optimal global fuel saving value (6.67 \%).

\section{Conclusions}
In this paper, we proposed an MPC-based solution for the energy management of a parallel hybrid tractor, able to deal with the traditional speed tracking requirement of the vehicle. The effectiveness of the solution is tested on an experimentally-validated simulation environment, developed after an experimental campaign oriented to the identification of every vehicle component. Results showed significant fuel saving -- 12.6 \% in the considered transport driving-cycle and 5.40 \% in the agricultural one -- without losing speed tracking performance.
   
\section*{Acknowledgments}
The authors would like to thank Argo Tractors for supporting the research, in particular Giovanni Esposito and Diego Palmieri.
   
\bibliography{biblio}
\bibliographystyle{IEEEtran}
\vspace{-2.7em}
\begin{IEEEbiography}[{\includegraphics[width=1in,height=1.25in,clip,keepaspectratio]{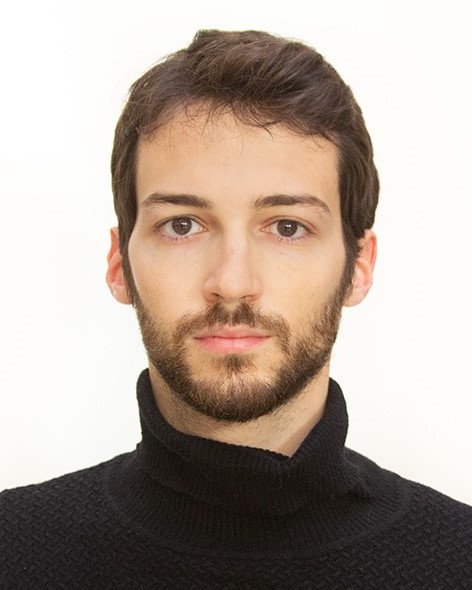}}]{Stefano Radrizzani} received B.Sc. and M.Sc. in Automation and Control Engineering from Politecnico di Milano, Milan, Italy, in 2017 and 2019, respectively, discussing a thesis about the braking pressure control of a brake-by-wire actuator for a Formula E. In July 2019, he was selected for a summer camp for elite students by the Hong Kong University of Science and Technology (HKUST), Hong Kong. In November 2019, he joined the mOve research group at Politecnico di Milano as a PhD candidate in Systems and Control. After a post-doc at The University of Alabama, Tuscaloosa, US, he is now a researcher at Politecnico di Milano. His current research is related to energy management, battery sizing and vehicles dynamics control in vehicles.
\end{IEEEbiography}
\vspace{-2.7em}
\begin{IEEEbiography}[{\includegraphics[width=1in,height=1.25in,clip,keepaspectratio]{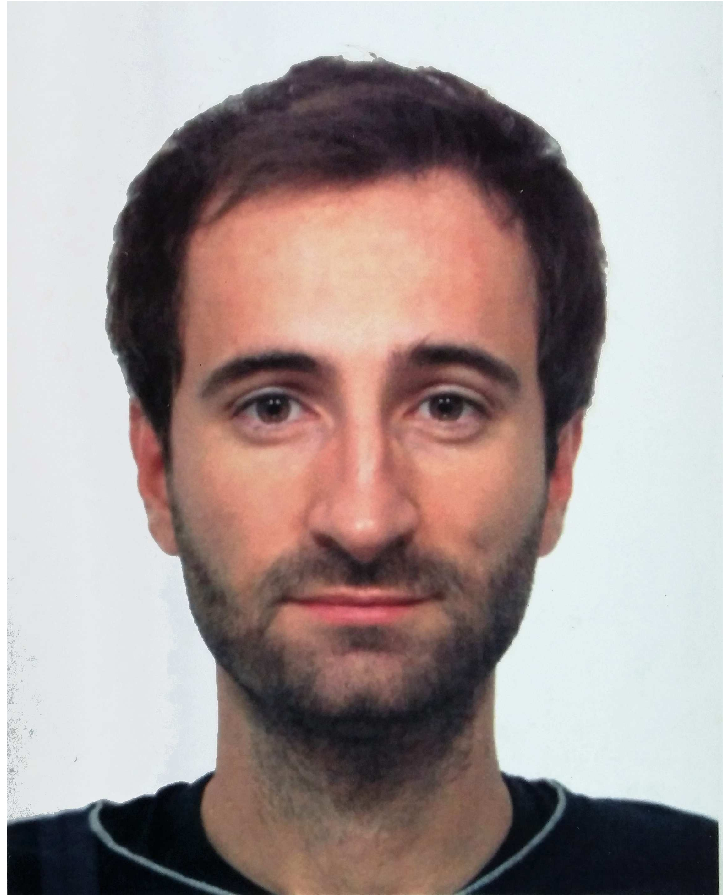}}]{Giulio Panzani} received a M.Sc in Mechanical Engineering in 2008 and a Ph.D. in Information engineering (system
control specialization) in 2012 from Politecnico di Milano, Milan, Italy. He held a post-doc position at the University of
Trento, Trento, Italy, and at the Swiss Federal Institute of Technology of {Z\"urich} (ETHZ), {Z\"urich}, Switzerland. He is an Associate Professor
with the Dipartimento di Elettronica, Informazione e
Bioingegneria, Politecnico di Milano, Italy. His main
research interests include the analysis of dynamics,
control design, actuation and estimation for two (and
four) wheeled vehicles.
\end{IEEEbiography}
\vspace{-2.7em}
\begin{IEEEbiography}[{\includegraphics[width=1in,height=1.25in,clip,keepaspectratio]{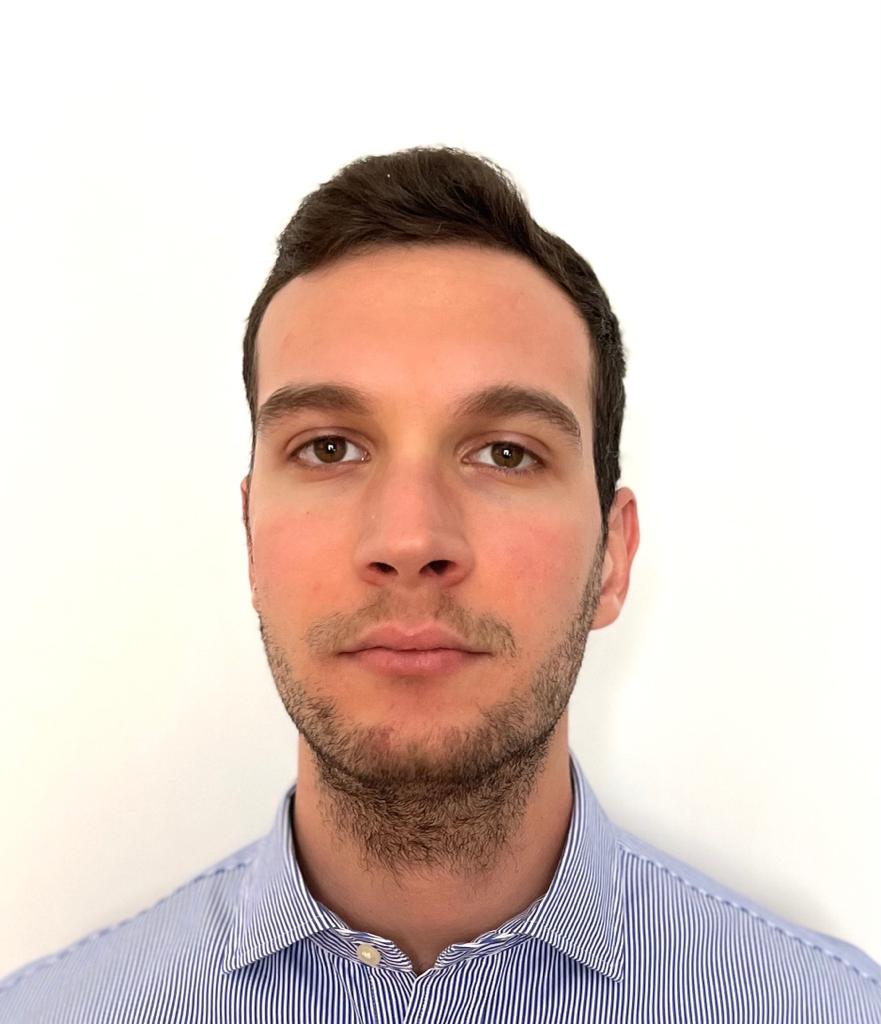}}]{Luca Trezza} received his Master’s degree in Automation and Control Engineering from Politecnico di Milano, Milan, Italy in 2021 discussing a thesis about the development of a model predictive control strategy applied to the Energy Management of a hybrid powertrain. In July 2021, he joined the mOve research group at Politecnico di Milano as Junior Research Assistant. His main interests are the control of semi-active dampers and, more generally, vehicle dynamics control strategies.
\end{IEEEbiography}
\vspace{-2.7em}
\begin{IEEEbiography}[{\includegraphics[width=1in,height=1.25in,clip,keepaspectratio]{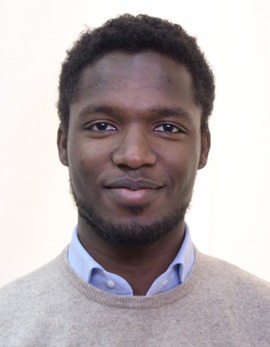}}]{Solomon Pizzocaro} received the B.Sc. on July 2016 and the M.Sc. on December 2018 in Automation and Control engineering from Politecnico di Milano, Milan, Italy. He defended a thesis concerning the autonomous navigation of a small two-wheeled vehicle for goods delivery in urban areas. From January to September 2019 he was a visiting scholar at the Clemson University International Center for Automotive Research (CU-ICAR), Clemson, SC, USA, were he participated in the Deep Orange 10 project, and worked on developing and implementing perception and navigation algorithms for a level 5 autonomous vehicle. He is currently a PhD candidate in System and Control at Politecnico di Milano and member of the mOve group. His main research topic is agricultural robotics.
\end{IEEEbiography}
\vspace{-2.7em}
\begin{IEEEbiography}[{\includegraphics[width=1in,height=1.25in,clip,keepaspectratio]{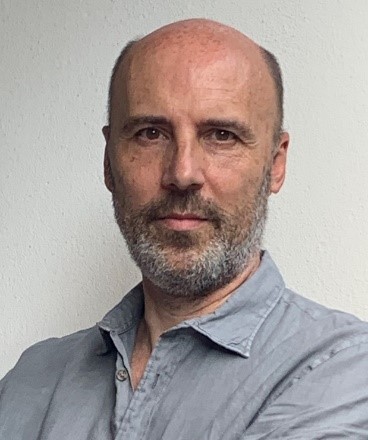}}]{Sergio M. Savaresi} received the M.Sc. degree in electrical engineering 1018
and the Ph.D. degree in systems and control engineering from the Politecnico di Milano, Milan, Italy, 
in 1992 and 1996, respectively, and the M.Sc. degree in applied mathematics from Catholic University, 
Brescia, Italy, in 2000. After the Ph.D. he worked as management consultant at McKinsey\&Co, Milan. He is Full Professor in Automatic Control at Politecnico di Milano since 2006 . He is Deputy Director and Chair of the Systems and Control Section of Department of Electronics, Computer Sciences and Bioengineering (DEIB), Politecnico di Milano. He is author of more than 500 scientific publications. His main interests are in the areas of vehicles control, automotive systems, data analysis and system identification, non-linear control theory, and control applications, with special focus on smart mobility. He has been manager and technical leader of more than 400 research projects in cooperation with private companies. He is co-founder of 9 high-tech startup companies.
\end{IEEEbiography}

\vfill\clearpage

\end{document}